\documentclass[prl,twocolumn,showpacs,preprintnumbers,floatfix,amsmath,amssymb]{revtex4}

\usepackage[dvips]{graphicx}
\usepackage{dcolumn}
\usepackage{bm}
\usepackage{color}

\begin{document}

{\color{red}Published in Phys. Rev. B {\bf 76}, 205437 (2007)}

\title{Quantum size effects in solitary wires of bismuth} 
\author{Shadyar Farhangfar} 
\email{shadyar@mpi-halle.de}
\affiliation{Department of Physics, University of Jyv{\"a}skyl{\"a}, FIN-40014 Jyv{\"a}skyl{\"a}, Finland} 
\affiliation{Max Planck Institute of Microstructure Physics, Weinberg 2, D-06120 Halle, Germany}
\date{\today}

\begin{abstract}
We have performed four-probe electrical  transport measurements on solitary highly crystalline wires of semimetallic bismuth with aspect ratios up to 60 at room and at cryogenic temperatures. By proper choice of the substrate material and the film deposition parameters, lithographic  wires with lateral dimensions of down to one single grain, $\sim 250$ nm, were fabricated.  The electrical resistance of each wire was measured against its thickness through successive reactive ion etching of the  self-same wire. Quantum size effects revealed themselves as regular oscillations in the electrical resistance. Some evidence for the semimetal-to-semiconductor  phase transition has been detected. The measured data are discussed within the framework of the existing theoretical models.  
\end{abstract}


\maketitle

Bismuth is a semimetal with unique  magnetic, electrical, and thermal properties. As prime examples, one can mention the large mean free path, large carrier mobility, small effective mass components of the charge carriers, and large magnetoresistance. These unusual properties arise from the highly anisotropic Fermi surface of this material. The small effective mass leads to a large de Broglie wavelength, $\sim40$ nm, making this element a potential choice for the experimental study of  quantum confinement effects. Furthermore, the exceptionally low magnetic susceptibility of bismuth and its very low thermal conductivity favor its use in  the investigation of certain other phenomena. While the former can be exploited for the study of magnetic impurity effects,  the latter is, in turn, of vital importance in thermoelectric applications.

In a microscopic semimetal, as the size of the sample shrinks, the overlaps between the valence and the conduction bands become smaller, evolving ultimately  to the formation of a semiconductor energy gap. Further reduction of the band overlaps will lead to a semiconductor-to-insulator phase transition \cite{ogrin66,sandom67}.  In principle, a whole quantum circuit, consisting of (semi)metal, semiconductor,  and insulator components, can be made just by tailoring the energy bands  in a bismuthlike semimetal.

Experimental studies of the quantum size effects (QSEs) in two-dimensional (2D) films of bismuth can be traced back to the mid 1960s \cite{ogrin66},  followed by a theoretical model to describe the measured data soon after \cite{sandom67}. Since then, numerous investigations have been performed to discover transport properties of bismuth at low dimensions. While study of the 2D films has been a subject of steady interest \cite{ogrin66}-\cite{rogaAPL03}, the one-dimensional (1D) structures have been tackled only recently \cite{zhang98}-\cite{farhangfar06}. Yet, despite all the advances in microfabrication techniques, experimental evidence for the observation of the QSEs in  the 1D regime has been scarce. This is mostly due to the lack of satisfaction of at least one of the following  necessary conditions: the lateral dimensions of the sample $w$ and $t$ have to be  comparable in size to the carrier wavelength $\lambda$;  the energy level spacing $\delta{E}$ has to be larger than the thermal broadening $k_{\rm B}T$, and it has to meet the uncertainty principle, $\delta{E}\gtrsim\hbar/\tau$ ($\tau$ is the relaxation time);  the surface roughness of the wire $\eta$ is to be smaller than $\lambda$; and  the fabricated wires should have  comparable band structures and morphologies. In addition, the wires have to be contacted  in such a way as to avoid  screening effects arising from the formation of oxide  or Schottky barriers. In what follows, we will first discuss the accomplishment of these conditions in our experiments. We will then introduce our measured oscillations of the electrical resistance as a manifestation of the quantum size effects. Throughout the paper,  theoretical arguments  will be employed to elucidate the experimental data.

   
The samples were fabricated by electron beam lithography and thermal evaporation techniques. To obtain wires with the desired quality, {\it i.e.} wires with large grain size, low carrier concentration, and small surface roughness, we examined the suitability of different substrates and deposition parameters. Figure 1 illustrates the dependence of the carrier concentration $n$ and the electrical resistivity $\rho$ on the substrate material and its temperature. The carrier densities  were obtained from Hall measurements  performed at liquid helium temperature  on  macroscopic samples grown under similar conditions. The extracted  resistivities have to be compared to  the values obtained for high-quality polycrystalline, $0.5\times10^{-5}\Omega$m, and  single-crystalline, $4\times10^{-5}\Omega$m, films \cite{yang99}. Wires with the largest grain sizes (and lowest carrier concentration) were obtained through the evaporation of bismuth at a rate smaller than 1 $\rm{\AA}s^{-1}$  and at a vacuum pressure of  better than  $10^{-6}$mbar onto a mica substrate kept at $90^\circ$C. The films grown on mica substrate were n-type, {\it i.e.} electrons were the majority charge carriers. The optimal deposition temperature  is in accordance with the recommended  range for the growth of  crystalline films, $\frac{1}{3}T_{\rm m}<T_{\rm s}<\frac{2}{3}T_{\rm m}$, where $T_{\rm s}$ is the substrate temperature and $T_{\rm m}$ is the melting temperature of the material to be deposited  \cite{komnik67}. Prior to the film deposition, to provide a cleaner interface for the growth, the resist and developer remnants in the exposed areas were removed in  an oxygen plasma ambiance.  Figure 2 shows an atomic force microscope (AFM) image of the wire segment so obtained. The constituent grains have a typical feature size of  about 200$-$300 nm.  Initially, prior to the first electrical characterizations,  the samples were further analyzed under AFM. While the asperity  is of  the order of $10$ nm (cf. the carrier wavelength) for the films deposited onto a mica substrate in the aforementioned conditions, it can often be much larger  for those grown onto the other substrates and/or in different conditions. Displayed in Fig. 3 is the x-ray diffraction (XRD) pattern of a  bismuth film grown onto a mica substrate. Clearly, the films are highly crystalline and have a preferred growth orientation. It would also be desirable to have transmission electron microscope (TEM) images or scattering patterns of the films as well; however preparation of the samples grown on mica for TEM characterization  proved problematic  \cite{sc}.        
\begin{figure}\label{Fig.1}
\begin{center}
\includegraphics[width=82mm] {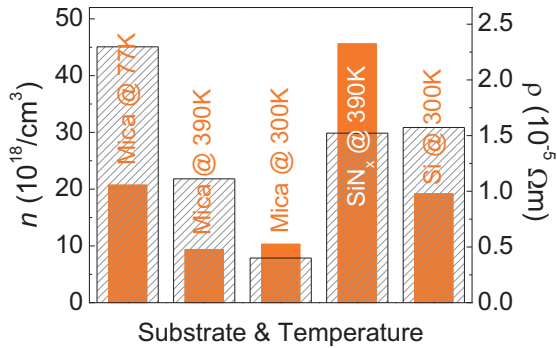}
\caption{(Color online) Dependence of the  carrier density $n$ (solid histograms) and the resistivity $\rho$ (dashed histograms) on the substrate material (mica, silicon, or silicon nitride) and  its temperature.}  
\end{center}
\end{figure}
It is also worth mentioning,  that despite all our efforts  (all together more than 100 samples went through the metalization procedure), the narrowest wires were more than 200 nm wide [full width at half maximum,  (FWHM)]. The main obstacle here was charging of the mica substrate during the e-beam lithography process,  which in turn led to the backscattering of the electron beam and, consequently, broadening of the exposed areas. Nevertheless, that small fraction of the wires with narrower widths subsequently failed to survive through the lift-off process or the following etching steps. However, as shown in Fig. 4, the calculated  density of states $D(E)$ for a 1D bismuth wire with dimensions comparable to those in the present study  suggests  feasibility of the detection of  QSEs in wires as wide as 300 nm. The level spacings here are in the range 1$-$10 meV, being an order of magnitude larger than the thermal broadening, $k_{\rm B}T\sim{0.4}$ meV, at liquid helium temperature.    
\begin{figure}\label{Fig.2}
\begin{center}
\includegraphics[width=82mm] {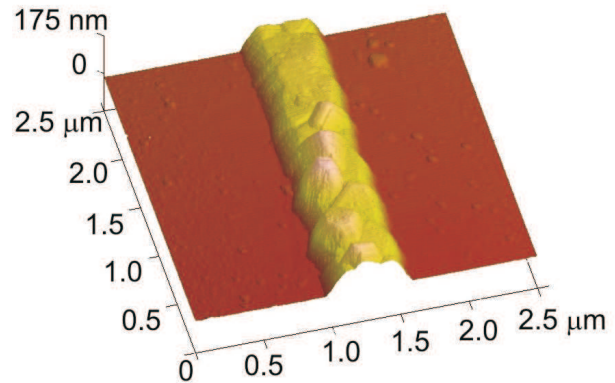}
\caption{(Color online) Atomic force microscope image of a bismuth wire segment deposited onto a mica substrate. The substrate temperature was $90^{\circ}$C.  The grains have a feature size larger than 200 nm. The wires have trapezoidal cross sections with sharply rising sidewalls.}
\end{center}
\end{figure}

Formation of  oxide or Schottky barriers  between the wires and the contact leads was circumvented by reducing the film deposition procedure to a single step.  In fact, observation of the low-bias plateaus in the current-voltage ($IV$) characteristics of the recent experiments \cite{croninThesis,TianAPL04} may be attributed to the existence of such barriers.             
Figure 5 depicts the $IV-$characteristic of a sample measured in liquid helium. The curve is linear down to the lowest bias currents.  Subsequent to the film deposition and the lift-off, each sample was measured at room, at liquid nitrogen, and at liquid helium  temperatures. All the measurements were performed in four-probe configuration, most of them through both the ac and the dc techniques.  To contact the samples to the measuring setup, we used conductive carbon paste. This material makes a robust contact to the pads and has a negligible contact resistance. Due attention was paid not to heat the samples by large currents; typical injection currents were in the range of 1$-$20 nA. 
\begin{figure}\label{Fig.3}
\begin{center} 
\includegraphics[width=82mm]{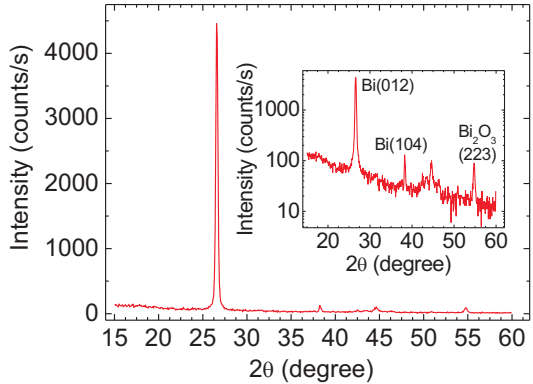}
\caption{(Color online) X-ray diffraction pattern of a bismuth film grown on a mica substrate. The film has a well-defined growth orientation (012). To make the smaller peaks visible, the vertical axis in the inset is presented on a logarithmic scale.  The Miller indices corresponding to the different orientations are taken from the JCPDS data base (codes 05-0519 for the Bi and 29-0236 for Bi{$_2$}O{$_3$}). The third peak from the left was not identified.}
\end{center}
\end{figure}
The reference frequencies in the ac-measurements were 10$-$20 Hz. After each series of measurements at the specified  temperatures, the sample was made thinner in a reactive ion etching  machine. To reduce the sample thickness by about 5$-$8 nm,  etching was done for a duration of 30 s with a 50 mW of argon plasma power. The temperature during this process was 30$^\circ$C. Our samples were typically 100$-$150 nm thick, $\sim300$ nm wide (FWHM), and 10$-$20 $\mu$m long. The electrical resistances of such wires were in the range 1$-$10 k$\Omega$ at room temperature. Most of the wires exhibited a resistance increase of up to even 100$\%$ at $T\approx77$ K followed by a further increase of about  10$-$30$\%$ at $T\approx4.2$ K. In contrast to this semiconductorlike behavior, occassionally  there were also samples reminicent of  the conventional metals. In what follows, we will consider samples of the former type.

Starting with an initial thickness of 100$-$150 nm, a series of  typically 8$-$15 measurements were performed until the sample resistance grew exponentially. (Indeed, in  most cases the samples were broken prior to this abrupt resistance increase.)  Before proceeding to a more detailed discussion of the measured data in the following, it would be illuminating to exclude the possibility that the observed oscillations could have stemmed from stochastic variations in the cross sections of the wires and the corresponding fluctuations in the electrical resistances. The resistance of the wire at the $i$th etching step can be written as ${R_i}\approx{\rho_i}l/[w({t_i}\pm{\eta_i})]$. Here, $l$, $w$, and $t_i$ stand for the length, width, and the thickness of wire respectively; $\eta_i$ is the asperity,  and $\rho_i$ is the corresponding electrical resistivity. The alteration in the resistivity of wire as its surface roughness changes can be estimated through the expression    $\rho_{i}^{-1}\sim\rho_{0}^{-1}\left[1-\frac{3}{16}(1-p_{i})\frac{{\Lambda_0}P_{i}}{S_{i}}\right]$, where $P_i$ is the perimeter, $S_i$ is the cross-section, and $p_i$ is the surface polish (a measure of the surface smoothness at the $i$th etching step; $p_{i}=1$ for the specular reflection of the carriers from the surface).  Furthermore, $\rho_0$ refers to the resistivity of a wire with perfect surface reflectivity and $\Lambda_0$ is the corresponding carrier mean free path \cite{ziman}.  Accordingly, within the course of two successive etching steps, the resistivity of wire remains almost intact, $\rho_{i+1}\approx\rho_{i}$. In  particular, in the beginning of the etching process, for a relatively smooth wire surface, one has $\rho_{i}\sim\rho_{0}$. It is now straightforward to show that the necessary condition for the electrical resistance to drop in a subsequent step, {\it i.e.} to have $R_{i+1}\lesssim R_{i}$, is that the average thickness removal between the two steps $\left|{\Delta {t_i}}\right|$ is smaller than the difference between their corresponding asperities, $\left|{\Delta {t_i}}\right|\lesssim\pm({\eta_{i+1}-\eta_{i}})$, a  condition that can hardly be satisfied. Nonetheless, the stochastic variations in the wire cross section can barely reveal themselves as regular oscillations with a clear periodicity in the thickness.

Figure 6(a) shows measured oscillations of the electrical resistance as a function of the wire thickness. The sample was originally $14$ $\mu$m long, $320$ nm wide, and $100$ nm thick. Apparently, there is a sharp enhancement in the resistance of the sample at around the tenth etching step, corresponding to a thickness of about 20$-$50 nm.  At this stage, we looked at another sample of the same batch under AFM. The film was  40$\pm$20 nm thick. This suggests an average removal of about 7  nm per etching step.  Subsequent to the 12th  etching step (not shown in the figure), the sample resistance increased to $\sim100$ k$\Omega$.   
\begin{figure}\label{Fig.4}
\begin{center}
\includegraphics[width=81mm] {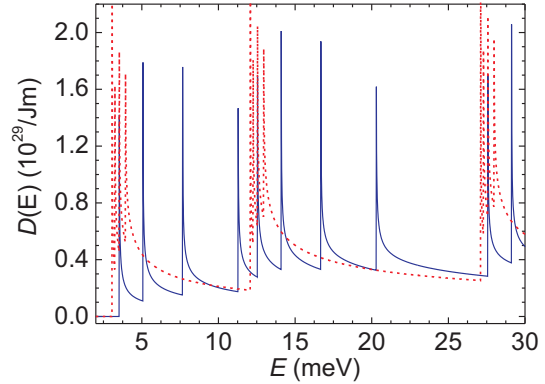}
\caption{(Color online) Calculated density of states $D(E)$ for a one-dimensional ($l\gg w,t$) bismuth wire  with lateral dimensions  $w=300$ nm, $t=50$ nm (solid blue line), and $t=150$ nm (dashed red line).   The wire was aligned along the bisectrix  axis. The  values for effective masses $m_x=0.00139$, $m_y=0.291$, and $m_z=0.0071$ (in units of the free electron mass) are taken from Ref. \cite{isaacson69}. For clarity, only the contribution of the few first subbands is shown. }
\end{center}
\end{figure}
This might point to the widening of the semiconductor energy gap already opened at around the tenth step. On the other hand, according to the theoretical models \cite{sandom67,farhangfar06}, the semimetal-to-semiconductor (SM-SC) transition thickness should equal the periodicity of the oscillations, {\it i.e.} $\sim14$ nm here.  A closer look at this issue would be in order. At low temperatures, the expression for the electrical conductivity in the SC regime  \cite{farhangfar06} can be estimated as         
\begin{equation}
\nonumber 
{{R_0}{w_0}{t_0}}\approx{R{w}{t}}\exp{\left(-{E}_{\rm g}/2k_{\rm B}T\right)},
\end{equation}
where ${E}_{\rm g}=E_{\rm g}(w,t)$  is the  energy gap.  $R_0$, $w_0$, and $t_0$ are the electrical resistance, the width and the thickness at the SM-SC transition.  $R$  and $t$  correspond to any arbitrary points on the  resistance-thickness plot, subsequent to the transition to the SC regime. Assuming $w\approx{w_0}$  and substituting  the measured values of parameters  [Fig. 6(a)] in the equation above, $R_0\approx13$ k$\Omega$, $t_0\approx 30$ nm, $R\approx22$ k$\Omega$ and $t\approx 23$ nm,  one obtains $E_{\rm g}\approx0.5k_{\rm B}T$.  This is comparable to what one  would anticipate  in the vicinity of the SM-SC transition, $E_{\rm g}\sim k_{\rm B}T$.    
\begin{figure}\label{Fig.5}
\begin{center}
\includegraphics[width=82mm] {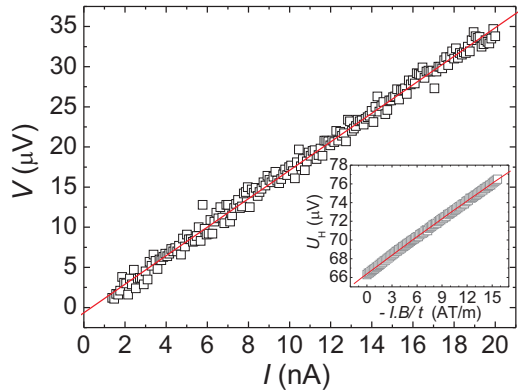} 
\caption{(Color online) $IV$ characteristic of a solitary bismuth wire with $w\approx300$ nm, $t\approx150$ nm, and $l=14$ $\mu$m   at $T\approx 4.2$ K. Note the linearity. (Inset)  Dependence of the Hall voltage $U_{\rm H}$ on the magnetic flux density $B$. Here $I$ is the current passing through the macroscopic Hall slab and $t$ is the thickness. Solid red line is a linear fit. The slope of the line corresponds to $-1/nq$ with  $n$ and $q$ being the  volume density and the charge of the carriers, respectively.}   
\end{center}
\end{figure}  

The measured data for another sample with $w=280$ nm, $t=150$ nm, and $l=18$ $\mu$m are presented in Fig. 6(b). This sample was damaged after the 16{th} etching step prior to its characterization at $T\approx4.2$ K. The sample resistance starts exponential growth at around the 14{th} etching step, corresponding to a transition thickness of about $t_0\approx38-80$ nm. The energy gap can be evaluated as explained above. Now, with $R_0\approx100$ k$\Omega$,  $R\approx250$ k$\Omega$ (estimated based on the resistance value at 77 K), and $t\approx 22-70$ nm,  we obtain $E_{\rm g}\approx(0.7-1.6)k_{\rm B}T$. The puzzling observation in Fig. 6(b) is the systematic decrease of the oscillation amplitudes at lower temperatures. A possible reason could be the formation of thermal stresses within the sample  which, in turn, might affect the morphology and the band structure. A thorough explanation, however, is obviously needed. 
\begin{figure}\label{Fig.6}
\begin{center} 
\includegraphics[width=82mm] {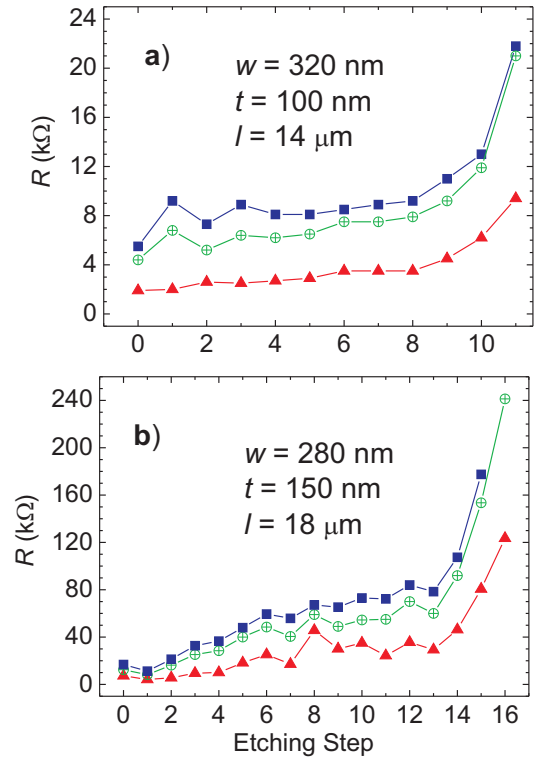}
\caption{(Color online) Dependence of the electrical resistance  of a bismuth wire on its thickness at liquid helium (solid squares), liquid nitrogen (crossed circles), and room (solid triangles) temperatures.  {\bf (a)} The sample was originally 320 nm wide, 100 nm thick, and 14 $\mu$m long.  Each etching step corresponds to a removal of about 5{$-$}8 nm film thickness.  {\bf (b)} Note the unexpected decrease of the oscillation amplitudes at lower temperatures. This sample broke after the 16th etching step, prior to its characterization in liquid helium.}
\end{center}
\end{figure}
The period of oscillations here,  as was the case with the previous sample, is about 10$-$16 nm. This is not surprising, given the fact that the samples were fabricated under similar conditions. However, the same disagreement between the values obtained from the threshold of the exponential tail and those estimated from the period of the oscillations persists. In addition, in both the samples, in contrast to the theoretical predictions \cite{sandom67, farhangfar06}, as the SM-SC transition is approached, the oscillations (at a fixed temperature) fade out.   These discrepancies  may be due to several reasons.  Apart from the experimental nonidealities ({\it e.g.}, the wires are not strictly single crystalline, but consist of an array of crystallites along the wire), in the existing theoretical models, the scattering potential $V_0$ as well as the density of states (DOS) were assumed to be homogeneous within the wire. In experiments, however, as noted above, the asperity $\eta$ of the sample grows abruptly at reduced dimensions and becomes comparable to the carrier wavelength, $\eta\sim\lambda$. As a consequence, the (grain) boundary scattering plays an increasingly important role and one has  $V_{0}=V_{0}(\eta;w,t)$. Similarly,  in a more sophisticated treatment, especially as the sample size shrinks, the DOS  should be substituted with the local density of states or with the spectral functions \cite{davies}. Furthermore, if the periods of oscillations have to become smaller as the specimens thin (as has been the  case in the experimental study of 2D bismuth films \cite{rogaAPL03}), the likeliness of missing the oscillations along the etching steps would increase substantially at lower thicknesses.  Also, as verified in the recent experiments \cite{crottini97}, the first layers of the grown films  may have  a different crystal structure. In  present study, this would mean alteration of the effective mass values as the film thins. Finally, let us notice that from the observed period of oscillations and the standing wave analogy, one can infer $\lambda\approx 32$ nm for the carrier wavelength and $n\approx2.5\times10^{17}$ cm$^{-3}$ for the volume density of the carriers.

  
To summarize, we have observed QSEs in solitary wires of bismuth. The n-type nanowires were only one single-crystallite wide. Suppression of the resistance oscillations as the wires thin points to  the dominance of surface scattering at reduced dimensions. Our findings could be helpful in further development of scalable solid-state devices comprised of band-engineered semimetals.

We would like to thank A. Schubert for the XRD patterns and K. Arutyunov for his assistance in the Hall measurements.



\begin{thebibliography}{07}
\bibitem{sc} It is to be noted that the TEM analysis would give  information on only a few monolayers of the grown film  over a limited  surface area of the specimen. The XRD technique instead asks for a much larger film surface (to gain measurable signal) and consequently cannot be used for characterization of a solitary nanostructure. Thus, calling the wires "single crystalline" on the basis of  diffraction patterns, as is a common usage in the literature, may not always be unambiguous.
\end{thebibliography}
\end{document}